# Common compression factor and bulk modulus quiescence points of liquid cesium

M. H. Ghatee*, M. Bahadori

*(Department of Chemistry, Shiraz University, Shiraz 71454, Iran)*

Fax: +98 711 2280928
E-mail: ghatee@sun01.susc.ac.ir

Abstract

*In this paper, a recently developed analytical equation of state (EOS) is used to investigate the bulk modulus of compressed liquid cesium and to locate common compression factor and the bulk modulus quiescence point(s). This EOS is applied quick well to Na and Rb far from $T_c$. Bulk modulus of liquid cesium have two quiescence points, a sharp one in the range* $1100\,\mathrm{K} - 1500\,\mathrm{K}$ *and a diffused one in the range* $1600\,\mathrm{K} - 1900\,\mathrm{K}$. *Therefore, two types of liquid cesium metal may be identified with characteristic structure and interaction potential energy. It is a constant independent of temperature, however, some residual change is seen due to the change in the values of integral of pair correlation function as temperature is increased. Furthermore, it is related to the shape of the unit cell and the atomic size at equilibrium. Observation of distinct liquid in the metal-nonmetal transition range is compared with NMR studies and molecular dynamic results.*

## 1. Introduction

The most important application of the theory of atomic properties is that of determining the stable crystal and liquid structures for a particular element. At a simple stage, one may wish to specify the atomic volume and elastic constants and examine the energy of different structures and different thermodynamic states at that volume. In the case of metal, particularly three types of energies must be specified. The first one, which is purely electrostatic, gives the contribution of interaction between ions to the energy of the system. The second one comes from the interaction between electrons. The third type of energy is the electron energy as a function of crystal momentum. [1] Interpretation of experimental data with the model electronic structure will allow understanding the details of structure of metals through modeling of the dynamics of the metal atoms.[2,3]

Macroscopic elastic behavior is the fundamental distinction between solids and liquids. They are similar in that each resists change of volume. However, solids resist change of shape while liquids do not. The stress tensors are used to define the extent of volume change and deformation, and to study the thermodynamics of deformation.[4] Both moduli of compression and moduli of rigidity are involved in determination of free energy. Then one can use the thermodynamic relation to derive the stress tensor, which turns out to be the negative of the ratio of pressure to moduli of compression. For small deformation, pressure and relative change in volume are small, and thus the ratio of pressure to the stress tensor, the bulk modulus $B_m$, may be written in differential form

$$B_m = \rho\left(\frac{\partial P}{\partial \rho}\right)_T \qquad (1)$$

where, P is the pressure, $\rho = 1/V$ is the molar density, V is the molar volume, and T is the absolute temperature.

The common bulk modulus point, which was observed by Huang and O'Connell [5], is considered as new empirical regularity for normal liquids. In their studies [5], it has been observed that the bulk modulus, which is the reciprocal of isothermal compressibility $\kappa_T$, as a function of density at different temperatures approach and intersect each other at (or within 5% of) a characteristic density.

This discovery has found to be applicable to a large number of normal liquids and their mixtures. To interpret this discovery, it has been noted that the van der Waals equation of state does not lead to the intersection of the isotherm.[5,6] However, the statistical mechanical equation of state derived by Song and Mason [6,7] based on the perturbation theory of dense liquids, characterize the intersection point by the specific density in terms of a parametric function of size and shape effects. The success of the second method leads to the failure of the first one in that the constant b, the van der Waals covolume, is actually a temperature dependent parameter as derived by the statistical mechanical approach.[6,7] Also by the application of the Linear Regularity Isotherm (LIR) [8,9], common intersection points have been observed for dense molecular liquids and their mixtures.

The investigation of thermodynamic properties of liquid alkali metals is complex due to their conduction electrons of atomic metal, though, in spite of their peculiarities, it is a challenge to be studied on atomic basis. [10, 11] Liquid cesium has been subjected to molecular dynamic simulation for pair potential and bulk modulus [10], experimental structural determination [12], and experimental thermodynamic measurement over the whole liquid range.[13] These plus a theoretical model allow drawing a conclusions about the liquid behavior. Close to the critical temperature, the fluctuation in density becomes large. Furthermore, close to the critical temperature, the metallic character of liquid alkali metals changes to a nonmetallic kind. The coordination numbers of these metals drop linearly with temperature [12,14-16] and the nearest adjacent distance increases slightly. Electronic structures of these metals manifest thermodynamic properties and known to be responsible for those peculiar transformations.

When the linear isotherm regularity (LIR) [17], formulated for Lennard Jones (12-6) fluids, was applied to liquid alkali metals, isotherms deviate from the linear behavior as the critical temperature is approached. [18] A conclusion was that the assumption of the nearest adjacent interaction, applied in the derivation of the LIR, breaks down. [18,19] This has been attributed to the onset of localization and partial correlation of valence electron. Then, to cover this shortcoming, exp6 potential function [20] was used to model the long-range and the short-range interactions in the liquid alkali metals. By application of the resulting exp6 isotherm to liquid cesium [20], the linear feature persists well beyond the density of the metal-nonmetal transition region. The success of the exp6 isotherm has been mainly attributed to its well-estimation of the attraction energy [20], which increases as the number of polarizeable atoms (forming clusters) are increased in the expanded liquid metal.

In the present work, the common compression and the common bulk modulus intersection points of the compressed liquid cesium are investigated by using the recently developed equation of state [21]. The common points are related to a limited temperature range, where indeed the temperature gradients of attraction and repulsion potential parts of the characteristic potential function e.g., parameters of the equation of state, are seized to quiescence variations, hence the title of the paper.

## 2. The equation of state

The equation of state is given here, while the details can be found else where [21]. It has been derived in the form,

$$(Z-1)V^2 = C + B\left(\frac{1}{\rho}\right) \quad (2)$$

where, Z is the compression factor. The potential function used to derive eq (2) is softer at short range and introduces more attraction at long range than the Lennard-Jones (12-6) potential function.[21] By some approximation the parameters of the isotherm [21]

$$C = \frac{4N\varepsilon \acute{o}^{6}}{K_{bcc}^{6} RT}, \qquad B = -\frac{2N\varepsilon \acute{o}^{3}}{K_{bcc}^{3} RT} \tag{3}$$

where $N$ is the Avegadro's number, the hard sphere diameter $\sigma = (2)^{-1/3} r_m$, $r_m$ is the position of potential minimum, and $RT$ has its usual meaning. $K_{bcc} = \left(3\sqrt{3}/4N\right)^{1/3}$. It is a constant characteristics of the (body centered cubic) unit cell of cesium. The parameters C and B are both explicit functions of temperature, and are related to the repulsive and attractive part of the applied potential function, respectively. [21] Since the experimental PVT data was used to determine C and B, the potential well-depth $\varepsilon$ becomes the binding energy of a pair of atom in ensemble of N-2 other similar atoms. It should be emphasis that the applied potential function is self-consisted by adopting the attraction and repulsion parameters simply by using experimental PVT data.

## 3. Results and Discussion

We have applied eq (2) to liquid Na and Rb metals. The results are satisfactory at low temperatures where a metallic property is enhanced. However, the available data does not extent to the corresponding metal-nonmetal transition range, and thus a conclusion cannot be made, as it can be in the case of liquid Cs metal.

We have used PVT data of liquid cesium [13] (up to 600 bar and 2000 K) to construct the isotherm equation (2). Isotherms of $(Z-1)V^2$ versus $1/\rho^* = \rho_c/\rho$ have considered over the whole liquid range, with $\rho_c = 2934.32\ \mathrm{mol/m^3}$ being the critical density.[22] The plots and are shown (for the range $1100\,K - 1950\,K$ only) in Figure 1 by thick lines. The perfect linear behavior of isotherms, found by plotting experimental data, over the whole liquid especially in the transition region is quite remarkable. (Linear behavior is quantified by $R^2 \geq 0.995$, where R is the linear correlation coefficient.) This feature is attributed to the characteristic of potential function used to derive the equation of state (2**).** [20,21] Notice that the transition to a nonmetallic fluid onsets at $T = 1350\,K$. Transition occurs at the density $9103.90\ \mathrm{mol/m^3}$ at any temperature in the range, indicating the influence of liquid structure on the transition.[12,19]

### 3. 1 The compression quiescence point

For the compression factor Z, at the density at which all isotherms intersect,

$$\left(\frac{\partial Z}{\partial T}\right)_{\rho_{Zq}} \approx \left[\frac{\partial (Z-1)V^2}{\partial T}\right]_{\rho_{Zq}} = 0 \tag{4}$$

where the subscript Z refers to compression factor and q refers to quiescence points which will be explained in section 3.3. To seek for the common compression point, we have extrapolated isotherms of equation (2) to higher densities. These are shown as thin lines in Figure 1. From Figure 1, we have hardly noticed that those isotherms belonging to the rage $1100\,K - 1500\,K$ all intersect each other at almost the same point (e.g., at $1/\rho^* = 0.208$, $\rho = 14107.31\ \mathrm{mol/m^3}$), and those belonging to higher temperature range ($1600\,K - 1900\,K$) are intending to intersect at a smaller value. However, the two common intersection points are so diffused that cannot be identified distinctively. (See section 3.2.) Notice that $T_c = 1938\ K$**.** It worth mentioning that the precise linearity of isotherm eq (2) at all temperatures, except some small wiggling near critical temperature, allows such precise extrapolations. The common point density is likely to be unique, however, the accuracy of intersection point is within 5%. As the temperature is increased, the intersection point moves slightly towards a higher density.

The density of quiescence point $\rho_{Zq}$ can be determined by

$$\left[\frac{\partial[(Z-1)V^2]}{\partial T}\right]_{\rho_{Zq}} = 0 = \frac{dC}{dT} + \frac{dB}{dT}\frac{1}{\rho_{Zq}} \tag{5}$$

and thus,

$$\rho_{Zq} = \frac{-(dB/dT)}{(dC/dT)} = \left(\frac{dB}{d(1/T)}\right) \Big/ \left(\frac{-dC}{d(1/T)}\right) \tag{6}$$

Equation (6) indicates that the density of the quiescence point is the ratio of temperature gradient of attraction and repulsion parameters.

Both C and B are quadratically smooth functions of T and linearly smooth functions of $(1/T)$. (See Figure 2a and 2b.) From Figures 2b, it can be seen that $\rho_{Zq}$ is basically independent of T. Therefore, according to eq (6), $\rho_{Zq}$ may be determined from the knowledge of temperature gradient of C and B at any temperature belonging to the liquid range of compressed cesium.

Expressions for parameters of eq (2) are related to molecular potential parameters [see eq (3)] and thus,

$$\rho_{Zq} = \frac{K_{bcc}^3}{r_m^3}\frac{[r_m(\partial\varepsilon/\partial T)_{r_m} + 3\varepsilon(\partial r_m/\partial T)_\varepsilon - \varepsilon r_m/T]}{[r_m(\partial\varepsilon/\partial T)_{r_m} + 6\varepsilon(\partial r_m/\partial T)_\varepsilon - \varepsilon r_m/T]}. \tag{7}$$

We have evaluated eq (7) by using values of $\varepsilon$ and $\sigma$ versus temperature [21], and by approximating $(\partial r_m/\partial T)_\varepsilon \approx (dr_m/dT)$ and $(\partial\varepsilon/\partial T)_{r_m} \approx (d\varepsilon/dT)$. A deviation plot of $\rho_{Zq}$ versus T with respect to average $\rho_{Zq}$ in the range $1100\,K - 1500\,K$ is shown in Figure 3. We can see that deviations in the range $400\,K - 1950\,K$ are between $-1\%$ to $9\%$, indicating that a sharp common intersection point is indeed valid only in a limited temperature range, and thus the intersection points are diffused and move towards higher densities as temperature is increased. It is interesting to see that the turn over in Figure 3 coincides with the temperature range where metal non-metal transition occurs. In addition, the turn over indicates that there might be two sets of isotherm with rather different intersection point. From the slopes before and after the turn over, it can be found that the accuracy for intersecting isotherms at a common point after the turn over is much less than the corresponding accuracy before turn over. This has also something to do with the nature of liquid Cs in the transition range and will be more characterized by studying bulk modulus. Notice that the position minimum of the turn over is not coincided exactly with the transition range, because we took the average only in the range 1100K-1500K.

Another feature that can be hypothesized by examining eq (7) is that $\rho_{Zq}$ is proportional to $K_{bcc}$ and $r_m$, where the former depends on the structure of the unit cell and the later depends on the size of the interatomic distance at equilibrium position. In other words, $\rho_{Zq}$ contains information about both the spatial orientation of the nearest adjacent atoms and the size of the interatomic distance. It worth noting that the feature of the expression for $\rho_{Zq}$ is the (slight) temperature dependence of $r_m$. We would expect that, in general, the value of $K_{bcc}$ to be changed with temperature. However, since primarily we have adopted $r = K_{bcc}\rho^{-1/3}$, then $K_{bcc}$ would become a constant of proportionality. [20,21] A similar role as for $K_{bcc}$ has been derived for non-polar fluids by using the ISM statistical mechanical equation of state in terms of $\lambda$, where $\lambda$ is known as the free parameter of the equation of state, and specifically as an arbitrary measure of molecular shape effect. [6]

### 3. 2. The bulk modulus quiescence points

Based on the perturbation method, Gubbins and O'Connell [23] have shown that

$$B_m^* = (\rho_n kT\kappa_T)^{-1} = \left[f_\kappa\left(\rho_n ó^3, kT/å\right)\right]^{-1} \quad (8)$$

where $\rho_n$ is the number density and k is the Boltzman constant. It was primarily assumed that for $\rho > 2\rho_c$, the dependence of $f_\kappa$ on the T is small and thus $B_m^* = \left[f_\kappa\left(\rho_n ó^3\right)\right]^{-1}$, which implies a law of corresponding states is applicable as far as different fluids can be describe by the same reference system. However, later Huang and O'Connell have been able to construct correlations based on eq (8). [5] (See also reference 24.). We need not to be concerned about the anisotropic contributions, because this has been of concerned in the derivation of eq (8) by using a reference system, defined by the orientational average of the full intermolecular potential, and may be applied to fluids with multipolar interaction provided that the anisotropy of the shape of the molecular core is small. In spit of these, Since Cs atom has zero dipole (and octapole) moment (but only non-zero quadarupole and hexadecapole moments), [25] we could assume that liquid cesium adheres reasonably well to this theory both at low and high densities.

Seeking for the common bulk modulus intersection point, the available experimental PVT data have to be extrapolated to lower (liquid) densities, and this has been done by solving the isotherm of equation of state eq (2) at particular temperature and density using an iterative procedure. Extrapolation of just the experimental PVT data for cesium [13] does not allow drawing such a conclusion. Hence, the equation of state (2) provides a valuable means for accurate extrapolation of PVT data too.

The eq (2) can be rearranged as a virial (like) equation of state in the form

$$Z = 1 + B\rho + C\rho^2. \quad (9)$$

Then $B_m^*$ would be in the form

$$1 - B_m^* = -2B\rho - 3C\rho^2. \quad (10)$$

Notice that at a given temperature, B and C are independent of density corresponding to the available pressure range. Differentiation of eq (10) with respect to temperature yields the density of the quiescent point

$$\rho_{Bq} = \frac{2}{3}\frac{-(dB/dT)}{(dC/dT)} = \frac{2}{3}\left(\frac{dB}{d(1/T)}\right)\Big/\left(\frac{-dC}{d(1/T)}\right). \quad (11)$$

where subscript B refers to the bulk modulus and q refers to quiescence (see section 3.3). It can be seen that $\rho_{Zq}^* = (3/2)\rho_{Bq}^*$, however, this relation is found true within 9.6%.

In Figure 4 two intersection points is noticeable, one belongs to the range $1100\,K - 1600\,K$ and other one belongs to the range $1700\,K - 1900\,K$ corresponding, respectively, to $1/\rho^* = 0.296$ $(\rho = 9913.23\,mol/m^3)$ and $1/\rho^* = 0.309$ $(\rho = 9496.17\,mol/m^3)$. The intersection in the later range is seen clearly in Figure 5. However, we can see the intersection point becomes diffused and moves towards lower $1/\rho^*$ as temperature is increased. The sharp quiescent point in Figure 4 is remarkable and that of Figure 5 is satisfactory. Although the common compression quiescent point at high temperature region is not sharp (section 3.1), studying the bulk modulus (Figure 4 and 5) distinctively confirm the existence of two different regions with characteristics liquid interaction and shape. This should be emphasized that although for the two common point's $1/\rho^*$ differ by 4.3% there is an appreciable vertical gap between them on $1 - B_m^*$ axis by about 24.5%, tracking into quite different trends.

### 3.3. The quiescence points

The main feature of $\rho_{Bq}$ is that its value is related to the high temperature region with the density about the metal-nonmetal transition density. Essentially, the results given above for $\rho_{B_q}$ correspond to expanded liquid range in which the temperature gradients of both repulsive and attractive coefficients, C and B, respectively are about to quiescence states (e.g., where curves are varying smoothly. See Figure 2a). Thus, $\rho_{Zq}$ and $\rho_{Bq}$ belong to the quiescence of both attraction and repulsion, hence the sub-subscript q of the corresponding densities. The general trend suggesting that a macroscopic evolution of parameters of eq (2) adequately capture the dynamic balance that exist between repulsive and attractive forces at (and around) the quiescent point.

The bulk modulus quiescent point $\rho_{Bq}$ in the low temperature region is remarkably sharp. Indeed, $\rho_{Bq}$ of a given isotherm moves slightly towards higher densities as temperature are increased. We attribute this feature to the change in structure of liquid, and to locate the position of common point reasonably, we apply the form of eq (10), which is appreciable in that it is equal to $1-\{1+\rho\int[g(r,T)-1]dr\}^{-1}$ where g(r,T) is the pair correlation function. [26] If a sharp common point could exist, then

$$\left[\frac{\partial\left(1-B_m^*\right)}{\partial T}\right]_{\rho_{Bq}} = \frac{\partial}{\partial T}\left\{1-\left[1+\rho\int\left(g(r,T)-1\right)dr\right]^{-1}\right\}=0\,. \qquad (12)$$

Therefore, it is required that the numerator of

$$\frac{-\left(\frac{\partial\rho}{\partial T}\right)\int\left(g(r)-1\right)dr-\rho_{Bq}\frac{\partial}{\partial T}\int\left(g(r)-1\right)dr}{\left[1+\rho_{Bq}\int\left(g(r)-1\right)dr\right]^2}, \qquad (13)$$

to be vanished. Otherwise the intersection point of a given isotherm with other isotherms intersecting at a nominal common quiescence point, say $\rho_{Bq}$, is subjected to a residual changes proportional to $\int[g(r,T)-1]dr$ and its temperature gradient.

In general $[g(r_{12})-1]$ is the total influence of molecule 1 on the molecule 2 located at distance $r_{12}$. Since the eq (12) is satisfied within a reasonable temperature range, two particular points can be discerned. In the first place, in a liquid system of N molecules, the total influence on molecule 1 by other N-1 similar molecules remains independent of temperature at a given density. In the second place, the residual change in expression (13) might be due to large (long-range) density fluctuation characteristic of low density expanded liquid cesium.

For cesium, graphical integration of $[g(r,T)-1]$ truncated at $r=14\,\dot{A}$ (e.g., $r/r_m=2.59$) as a function of temperature is shown in Figure 6. We have used plots of pair correlation function calculated by using the neutron scattering experiment.[12] As the temperature increases, $d\left(\int[g(r,T)-1]dr\right)/dT$ becomes smaller. In the range $1100\,K-1500\,K$ it turns over and then increases beyond 1500K. Examination of the two terms in the numerator (13) and the plot in Figure 6 shows that before the turn over there is possibility for $\left[d\left(1-B_m^*\right)/dT\right]$ to be exactly equal zero, however, beyond the turn over $\left[d\left(1-B_m^*\right)/dT\right]\neq 0$ and a residual change in the position of intersection is inevitable. These are consistent with a quite sharp intersection point in the range 1100K-1500K and consistent with a diffused one beyond 1500K (see Figures 4 and 5.). The turn over is actually due to change in the sign of $d\left(\int[g(r,T)-1]dr\right)/dT$ from negative to positive. Thus, the conclusion is that in the second region the structure of system is in a rather constant change as the temperature is increased. The change in structure could be resulted from a strain in the unit cell, which could lead to change in the coordination number. Belashchenkov *et al*., [10] in their molecular dynamic simulation of liquid cesium, has noticed differences between experimental bulk modulus and the corresponding molecular dynamics simulation $\Delta B_m$. However, as the metal expands $\Delta B_m$ decreases sharply and becomes smaller than the error of the experimental bulk modulus at $T_c$. Although the accuracy of their calculation is not so high, they have attributed their results to the contributions of the free electron in the metallic region, which is absence in (their) molecular dynamic simulation. Evidently, as liquid cesium metal expands, the free electrons are diminished in extent and electron-ion correlation increases. In spite of this prediction, the result of their experiment is idle to predict liquids with different bulk moduli. On the other hand Warren *et al*. [15], in a NMR investigation of the electronic structure of expanded liquid cesium have concluded that as the liquid density is reduced below about 1.6-1.4 gr/cm$^3$ (12038.22-10533.44 mol/m$^3$) liquid takes on unusual electronic character due to qualitative changes developed. The susceptibility enhancement increases sharply, Korringa relation breaks down, indicating a shift from ferromagnetism to antiferromagnetism enhancement and the wave function spread out so as to cause a substantial reduction of electron density at the nucleus has been concluded. This effect is reversed only at the lowest densities below 0.8 gr/cm$^3$ (6019.11 mol/m$^3$), roughly twice as the critical density. Examination of bulk modulus in this work indicates existence of two types of liquid cesium metal, corresponding to the highly correlated metal up to the metal-nonmetal transition region and the very low-density liquid. Our results, depicted in Figure 4, which show that there are gaps between the isotherms and between the corresponding common point of low and the very low density (expanded) liquid are in agreement with NMR studies. However, the behavior of $\Delta B_m$ resulted from molecular dynamics simulation [10] nicely indicate a transition from metallic character to a nonmetallic one when liquid Cs is expanded.

Application of pressure (or thermal increase of molecular motion at a given volume) forces a strain, which at the molecular level leads to deformation of the structure of the unit cell, though determination of the strained unit cell requires precise simulations.

Experimental PVT data of cesium has been updated by Kozhevanikov in 1991 [22], however, the number of data points, in the temperature range, are less than in reference 13. We have used also these data to check the method of this study, and have detected two distinct $1/\rho^*_{Zq}$ and $1/\rho^*_{Bq}$. Although, the sharpness of these characteristics density confirms capabilities of our method, their position are somewhat shifted with respect to those obtained by data of reference13. This may be attributed to the accuracy of the data, and to the differences in the number of data points which essentially affect the results of the extrapolations.

Most simulation procedures do not include thoroughly a direct contribution of extent of free electron to the thermodynamic properties. This study offers recommendation for computer simulation of thermodynamic properties by including a programmed electron correlation in the formation of clusters of different sizes, which may be able to predict bulk modulus of liquid metal, a typical one for liquid cesium depicted in Figure 4.

## 5. Conclusions

The characteristic equation of state for liquid cesium has been used to investigate and locate the common compression quiescence point and common bulk modulus quiescence points. Bulk moduli has two common quiescent points which, according to the equation of state, belong to the thermodynamic state at which the repulsion and attraction terms of the interaction potential function are seized to quiescence states. Densities at intersection points are independent of temperature. They are proportional to the ratio of temperature gradients of parameters of the equation of state C and B. The general trend suggesting a macroscopic evolution of the parameters of the equation of state adequately captures the dynamic balance that exist between repulsive and attractive forces at (and around) the quiescence point.

The densities of common bulk modulus quiescence points are almost within the range of available experimental data. Whereas, experimental data must be extended far to the high-density region to check for the common compression quiescence point. Direct exact measurements are required under rather moderate condition to prove such assertion for the case of bulk modulus.

The characteristic density at the common intersection point is valid over the whole liquid range within <10% and over the metal-nonmetal transition range within 1%. The intersection point is rather sharp in the metal-nonmetal transition range $1100\,\mathrm{K} - 1600\,\mathrm{K}$ but is rather spread in the range $1600\,\mathrm{K} - 1900\,\mathrm{K}$ and moves to higher density as the temperature is increased. Observation of two bulk modulus quiescence points are in accord with the NMR investigation but not with the molecular dynamic, suggesting for a quantum mechanical exploration of metallic system by the later method including the extent of the free electron contribution to the thermodynamic properties in the metallic range, nonmetallic range, and close to the critical point.

**Acknowledgements**

The authors thank the research committee of Shiraz University for supporting this project, grant 79-SC-1369-C123. The authors are also grateful to the referee for bringing reference 24 to our attention.

## Figure captions

**Figure 1**. Plots of $(Z-1)(V/V_c)^2$ versus $1/\rho^*$ of liquid cesium over the whole liquid range (shown the range 1100 K- 1950 K only). The solid lines belong to the available experimental PVT data. The thin lines are the related extrapolations.

**Figure 2a**. Plots of C and B versus T.

**Figure 2b**. Plots of C and B versus 1/T.

**Figure 3**. Plot of percent deviation of $\rho_{Zq}$ at all temperature (with respect to the averaging $\rho_{Zq}$ in the range $1100\,\text{K} - 1500\,\text{K}$).

**Figure 4**. Plots of reduced bulk modulus versus $1/\rho^*$ in the range $1100\,\text{K} < T < 1600\,\text{K}$ .

**Figure 5**. The same as Figure 4 except for the range $1700\,\text{K} < T < 1950\,\text{K}$ .

**Figure 6**. Graphical integration of $[g(r,T) - 1]$ for liquid cesium versus temperature

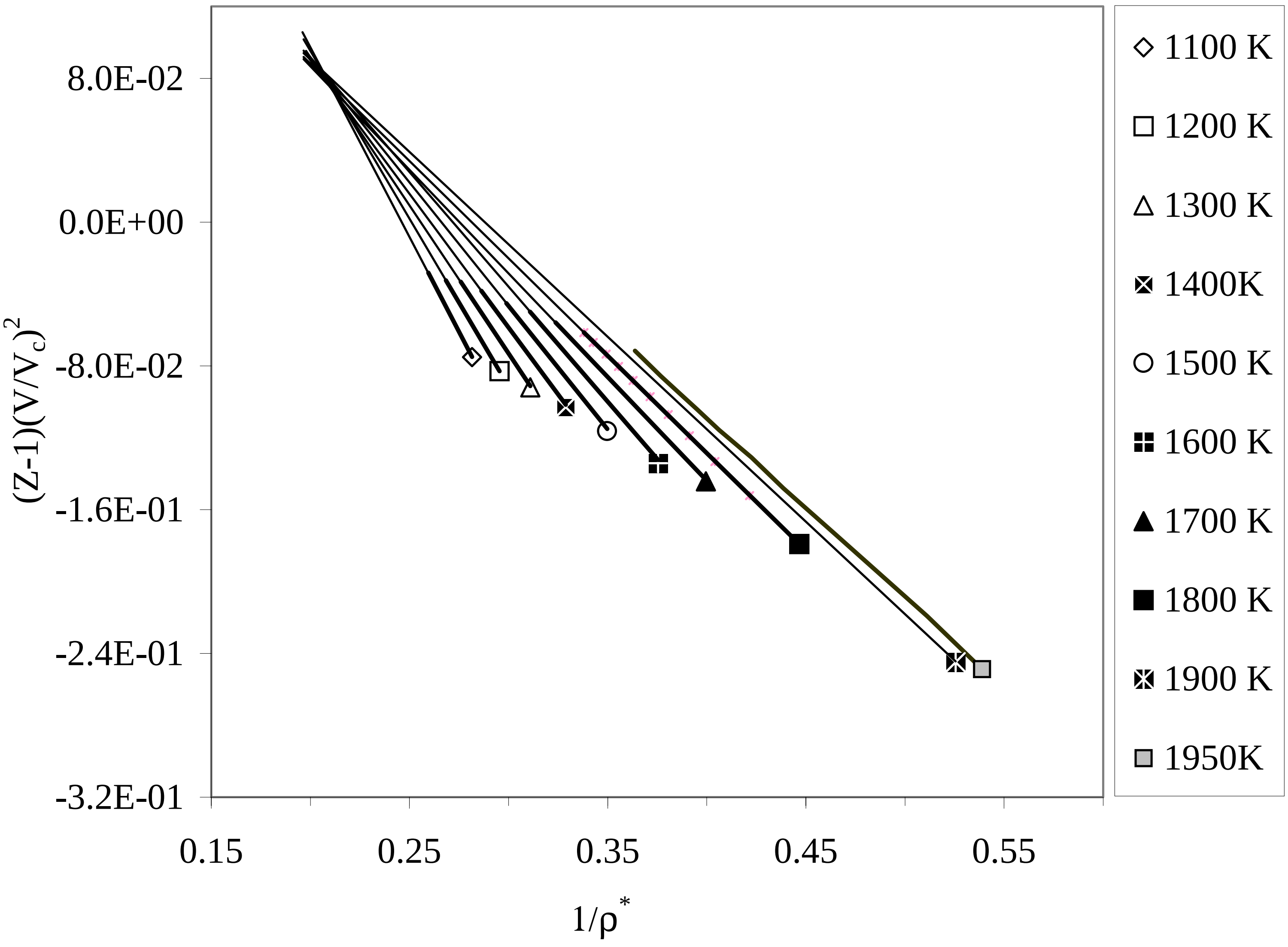

Figure 1

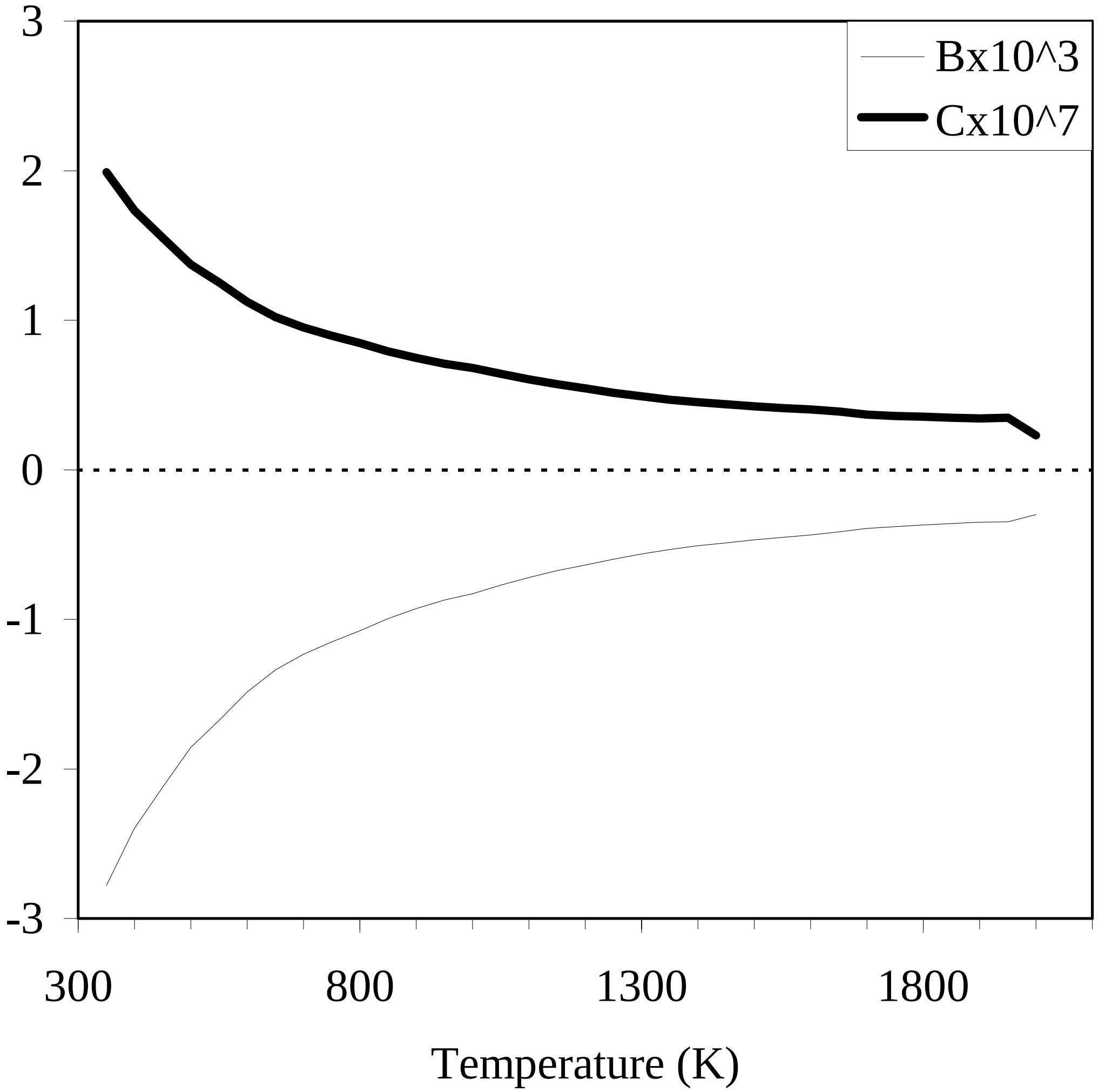

Figure 2a

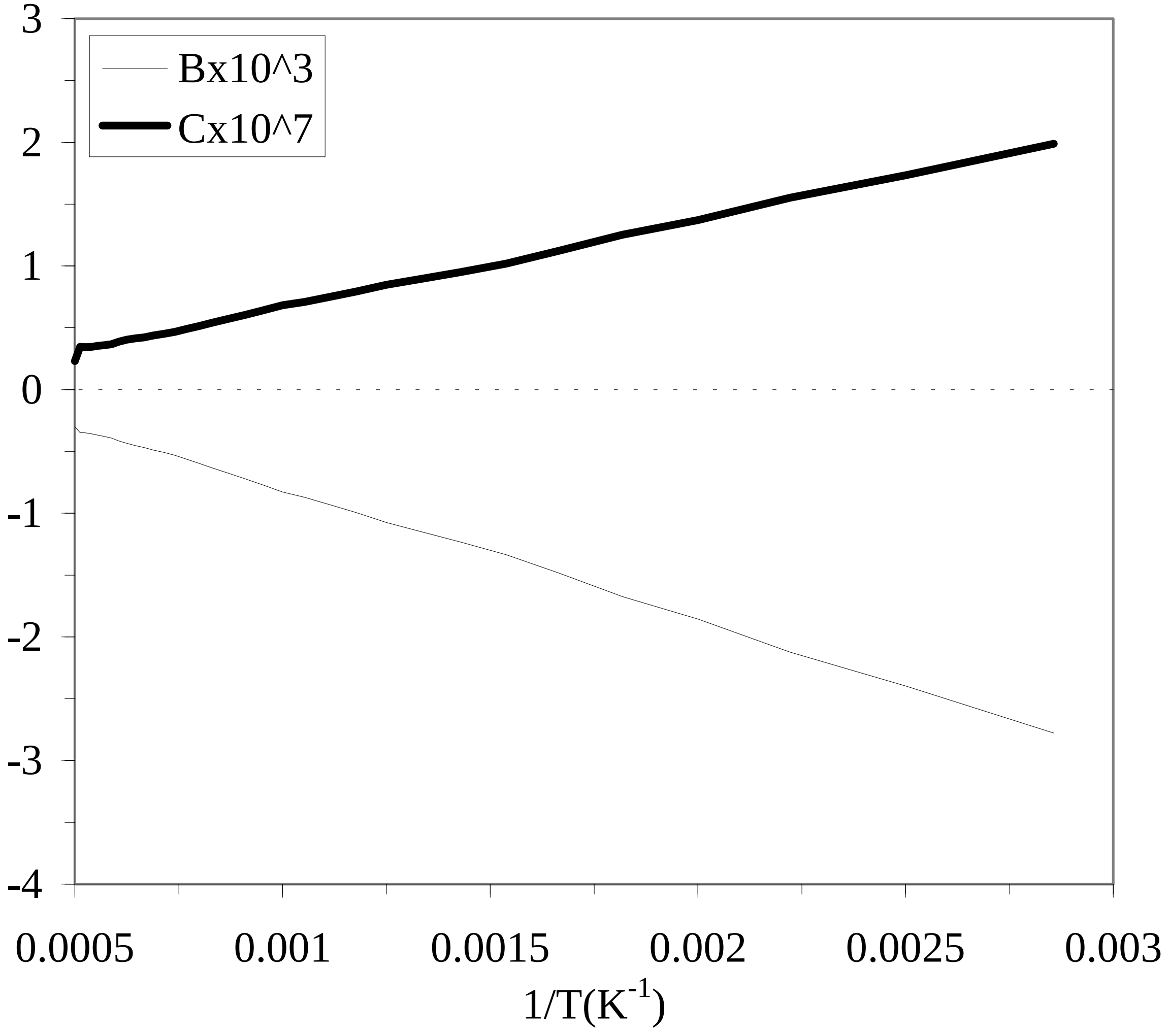

Figure 2b

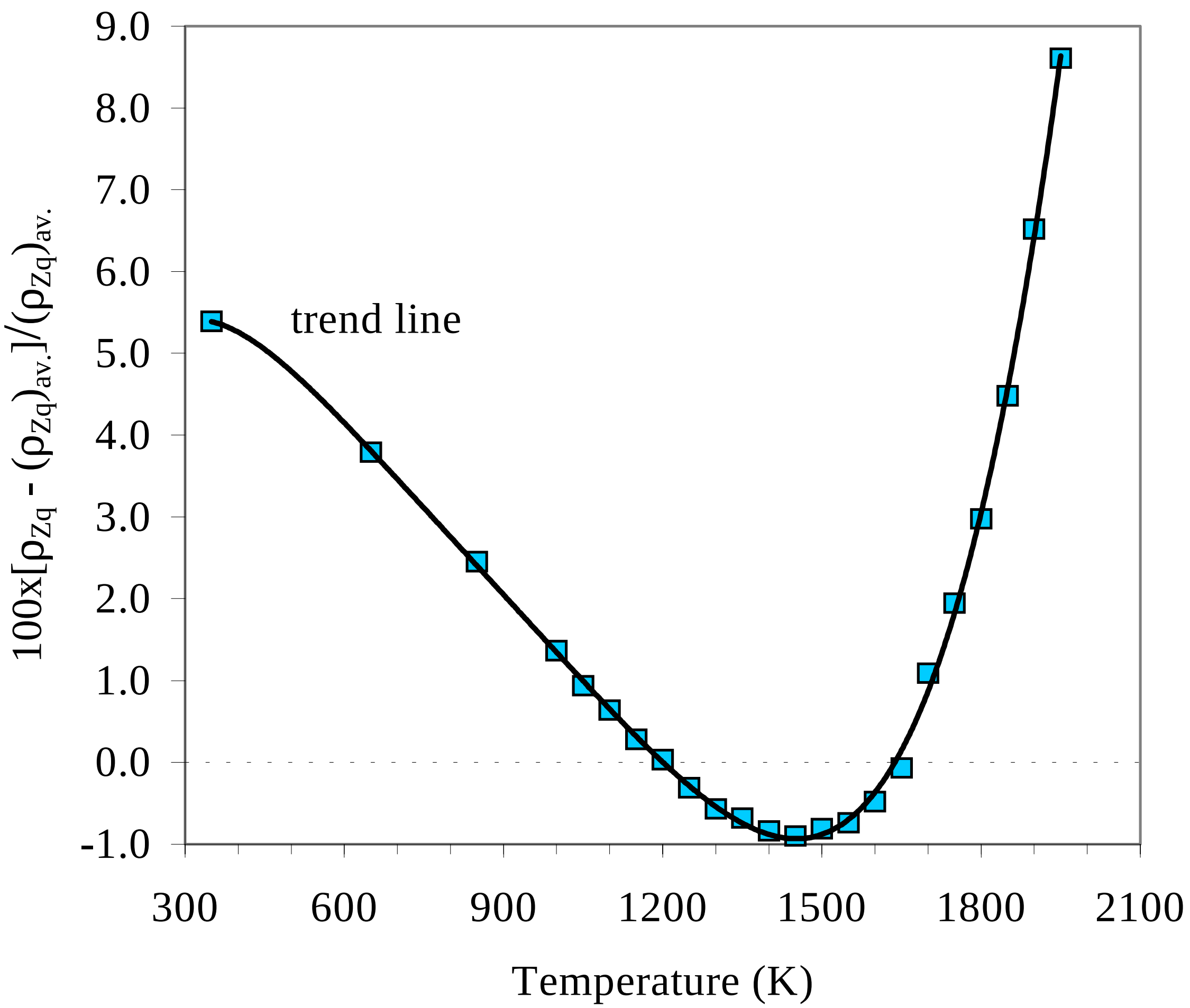

Figure 3

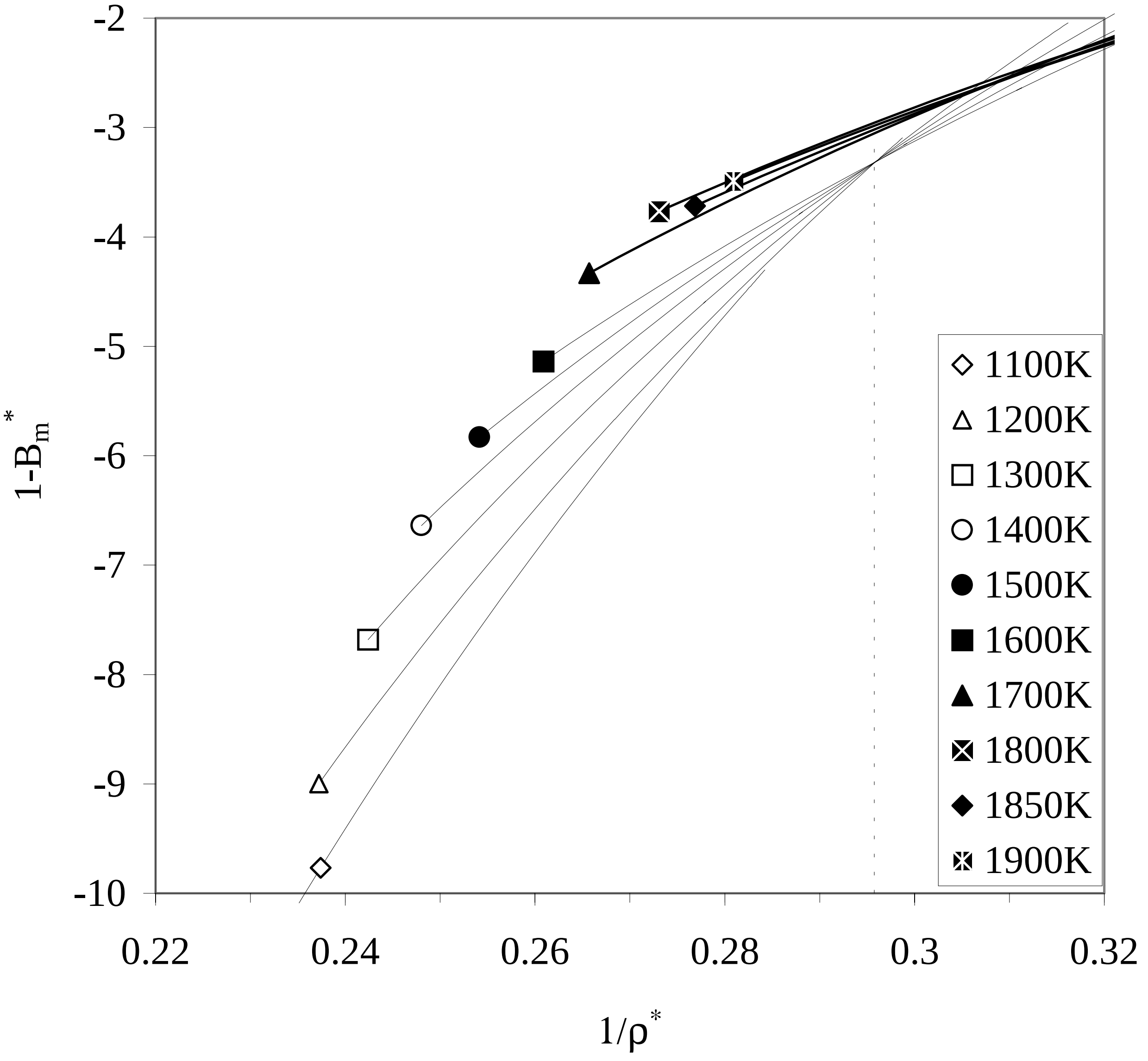

Figure 4

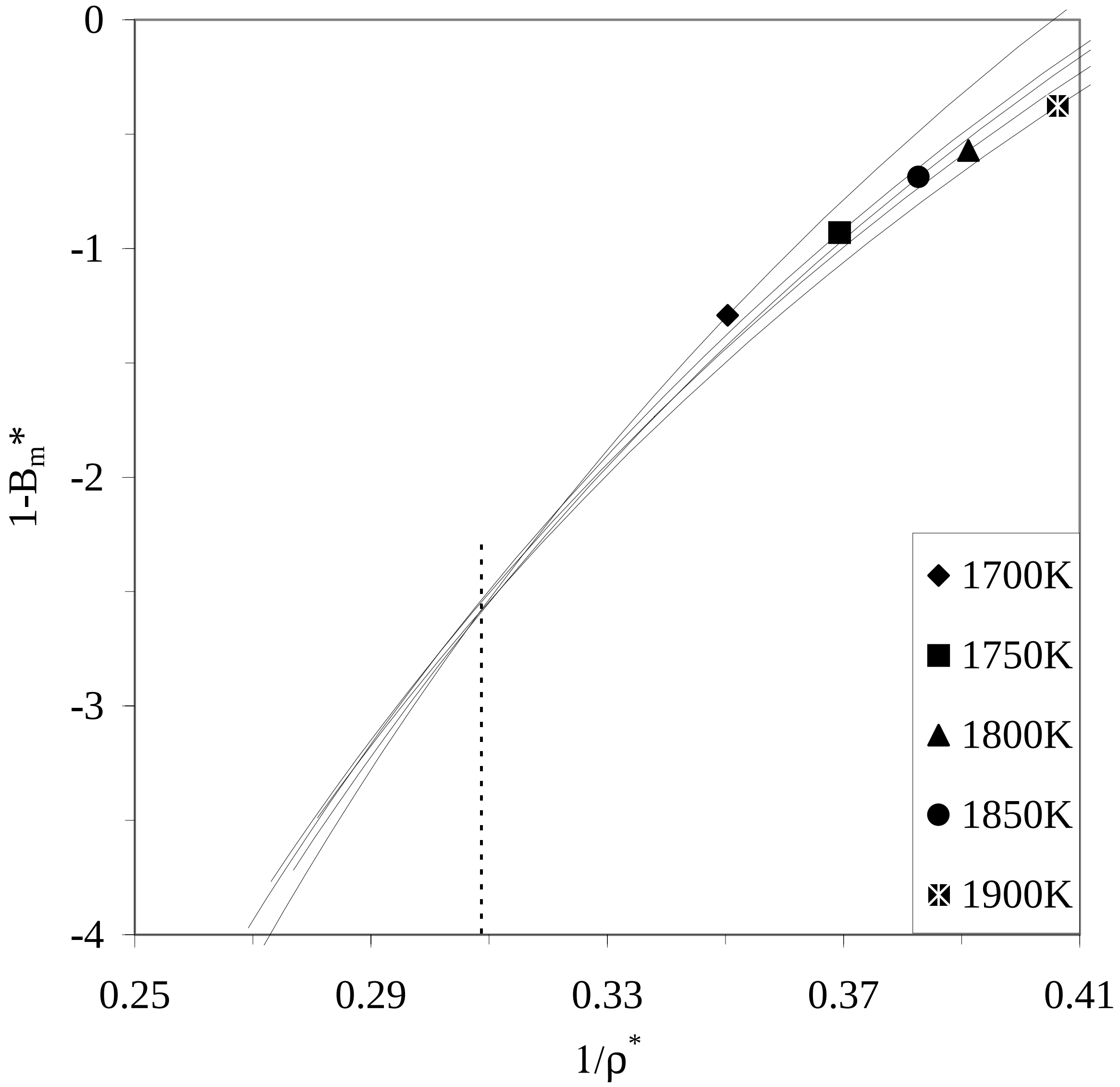

Figure 5

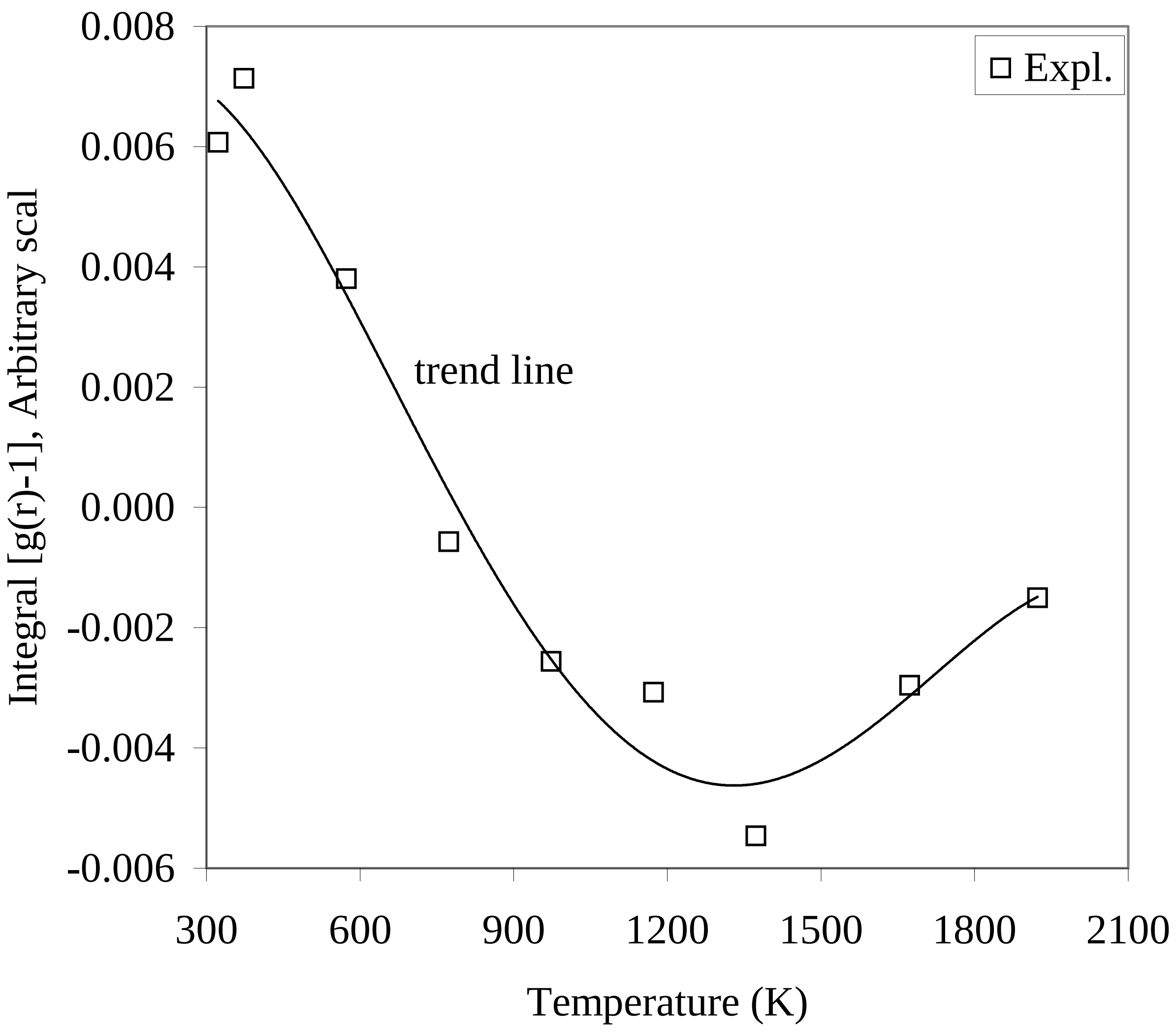

Figure 6